\documentstyle[12pt]{article}

\thicklines                         
\def\d{\delta}
\def\e{\epsilon}                
\def\f{\phi}                    
\def\g{\gamma}

\def\j{\psi}

\def\l{\lambda}
\def\m{\mu}
\def\n{\nu}

\def\p{\pi}                     
\def\s{\sigma}                  

\def\D{\Delta}

\def\J{\Psi}

\def\cd{{\cal D}}

\def\cg{{\cal G}}
\def\ch{{\cal H}}   

\def\co{{\cal O}}

\def\car{{\cal R}}

\def\cu{{\cal U}}

\def\cbo{{\,\raise-.15ex\Sc [\,}}                       
\def\dg{^\dagger}                                     

\def\Bar#1{\overline{#1}}                       
\def\bra#1{\Big\langle #1\Big|}                 
\def\ket#1{\Big| #1\Big\rangle}                 
\def\vev#1{\Big\langle #1 \Big\rangle}           
\def\abs#1{\Big| #1\Big|}                       

\def\Braket#1#2{\Big\langle #1 \Big| #2 \Big\rangle}
\def\sbra#1{\left\langle #1\right|}             
\def\sket#1{\left| #1\right\rangle}             
\def\svev#1{\left\langle #1\right\rangle}       

\def\ddt#1{{\buildrel {\hbox{\LARGE .\kern-2pt.}} \over {#1}}}

\def\secteq#1{ \setcounter{equation}{0}
               \renewcommand{\theequation}{#1.\arabic{equation}} }

\def\beqn#1{ \renewcommand{\theequation}{#1}
             \begin{eqnarray} }
\def\eeqn{ \renewcommand{\theequation}{\arabic{equation}}
           \end{eqnarray} }

\def\beqr#1{ \setcounter{equation}{#1}
             \begin{eqnarray} }

\def\eeqr{\end{eqnarray}}
\def\NON{\nonumber\\}
\def\beqrabc#1{ \setcounter{equation}{0}
                \renewcommand{\theequation}{#1\alph{equation}}
                \begin{eqnarray} }
\def\beqrn#1#2{ \setcounter{equation}{#2}
                \renewcommand{\theequation}{#1.\arabic{equation}}
                \begin{eqnarray} }
\def\seeq#1{eq.~(\ref{#1})}
\def\seEq#1{Eq.~(\ref{#1})}
\def\seeqs#1{eqs.~(\ref{#1})}

\def\seneq#1{~(\ref{#1})}
\def\rf{ref.~\cite}

\def\APH#1{Ann. Phys. {\bf #1}}
\def\CMP#1{Comm. Math. Phys. {\bf #1}}

\def\NPB#1{Nucl. Phys. {\bf B#1}}
\def\NPBP#1{Nucl. Phys. (Proc. Suppl.) {\bf B#1}}
\def\PLB#1{Phys. Lett. {\bf B#1}}
\def\PRD#1{Phys. Rev. {\bf D#1}}
\def\PR#1{Phys. Rev. {\bf #1}}
\def\PRL#1{Phys. Rev. Lett. {\bf #1}}

\def\sstyle{\scriptstyle}

\def\rhs{\mbox{r.h.s.} }
\def\lhs{\mbox{l.h.s.} }
\def\ie{\mbox{i.e.} }

\def\etc{\mbox{etc.} }

\def\frac#1#2{ {\sstyle {#1\over #2} } }
\def\det#1{{\rm det}\left(#1\right)}

\def\tr{{\rm tr}\,}

\def\half{{1\over 2}}
\def\Re{{\rm Re\,}}

\def\det{{\rm det\,}}

\def\sign{{\rm sign}}

\def\seff{S_{\rm eff}}
\def\seft{S^\infty_{\rm eff}}
\def\uar{\uparrow}
\def\dar{\downarrow}

\def\beq{\begin{equation}}
\def\eeq{\end{equation}}
\def\bqry{\begin{eqnarray}}
\def\eqry{\end{eqnarray}}

\begin{document}
\hyphenation{fer-mio-nic}
\noindent  \hfill WIS--94/19--May--PH

\noindent  \hfill TAUP--2148--94

\noindent  \hfill hep-lat/9405004
\par
\begin{center}
\vspace{15mm}
{\large\bf
Axial Symmetries in Lattice QCD with Kaplan Fermions}\\[10mm]
Vadim Furman\\
School of Physics and Astronomy\\
Beverly and Raymond Sackler Faculty of Exact Sciences\\
Tel-Aviv University, Ramat Aviv 69978, ISRAEL\\[5mm]
{\it and}\\[5mm]
Yigal Shamir\footnote{
\noindent Present address:
School of Physics and Astronomy,
Tel-Aviv University, Ramat Aviv 69978, ISRAEL}\\
Department of Physics\\
Weizmann Institute of Science, Rehovot 76100, ISRAEL\\[15mm]
{ABSTRACT}\\[2mm]
  \end{center}
\begin{quotation}
  A lattice definition of QCD based on chiral defect fermions is
discussed in detail. This formulation involves
(infinitely) many heavy regulator fields,
realized through the introduction of an unphysical fifth dimension.
It is proved that non-singlet axial symmetries become exact
in the limit of an infinite fifth dimension, and before the continuum
limit is taken.
\end{quotation}

\newpage
\thispagestyle{empty}
\mbox{  }
\setcounter{page}{1}

\newpage
\noindent {\large\bf 1.~~Introduction}
\vspace{3ex}
\secteq{1}

  Consistent local
regularization methods which preserve gauge invariance must break
axial symmetries explicitly.  This is a consequence of the well-known axial
anomaly~\cite{an}. However, flavour Non-Singlet Axial Symmetries (NSAS for
short) are recovered in renormalized correlation functions to all orders in
perturbation theory~\cite{ab,rg}.  (For recent progress and references to
earlier literature see \rf{mb}).

  Going beyond perturbation theory, the rigorous definition of QCD relies on
the lattice regulator. The most popular method to avoid the fermion doubling
problem employs Wilson's prescription for the fermion action~\cite{w}.  The
advantage of this method is in its simplicity. But the ensuing breaking of
axial symmetries is hard, in the sense that perturbative corrections to quark
masses are $O(1/a)$ where $a$ is the lattice spacing. Hence, one has to fine
tune the bare quark masses in order to recover the correct renormalized masses
in the continuum limit.

  Using Wilson fermions, it was shown that weak coupling perturbation theory
(WCPT) on the lattice reproduces the axial anomaly~\cite{ks}, and that NSAS are
recovered to all orders in WCPT in the continuum limit~\cite{ks,it}. These
results have in fact some validity beyond the scope of WCPT, and one can
discuss the renormalization of gauge invariant composite operators.  But,
because of the severe fine tuning problem inherent to Wilson fermions, one
cannot give a completely general non-perturbative proof of the restoration of
NSAS.

  {\it Assuming} the existence of the chiral limit in the full quantum theory,
what one can do in numerical simulations with Wilson fermions is to
determine the correct finely tuned values of the bare masses by  measuring
some correlation functions.  Fixing the bare parameters this way, one hopes
that NSAS will be recovered in all other correlation functions, thus
reproducing for example the results of current algebra~\cite{ca}.

  This situation is unsatisfactory for several reasons. On the theoretical
level, one would like to have a true non-perturbative proof of the restoration
of NSAS. Moreover, the construction of a lattice model of QCD where the
existence of the chiral limit can (a) be proved and (b) does not require any
fine tuning, is important for practical reasons.  To elucidate the
importance of such framework, we can mention for example the problems involved
in measuring weak matrix elements on the lattice with Wilson
fermions~\cite{weak,it}. The fine tuning problem is not over when the
bare masses have been fixed. Because of the hard breaking of the axial
symmetries, the definition of renormalized four fermion operators which
are necessary for the computation of weak decays, involves additional
fine-tuning.  Moreover, some of the relevant operator mixings receive genuinely
non-perturbative contributions, which cannot be determined even in principle
by short distance expansions such as the OPE.

  An alternative lattice formulation which does enjoy a certain
degree of axial symmetry is the staggered fermions formulation.
The staggered fermions action has a $U_V(1)\times U_A(1)$ symmetry,
and the exact $U_A(1)$ can be used for a better determination of meson decay
constants~\cite{weakstg}. However, this formulation also has several drawbacks.
The number of flavours must be equal to four, and disentangling  spacetime
and flavour symmetries is a non-trivial issue. Moreover, in the continuum limit
one expect to recover the full $SU(4)$  axial flavour symmetry, but on the
lattice only one of the corresponding currents is conserved. The other
fourteen currents presumably suffer from the same problems as in the case of
Wilson fermions.

  In this paper we present a new lattice formulation of QCD with a very mild
breaking of all non-singlet axial symmetries. The formulation is based on the
introduction of many (in the chiral limit infinitely many) heavy ``regulator''
fields~\cite{sf,k,nn1}. More specifically, we use a variant~\cite{bndr}
of Kaplan's proposal~\cite{k} to realize
light ordinary fermions as zero modes bound to four dimensional defects
in a theory of massive five dimensional Dirac fermions~[12-21].
The right-handed (RH) and left-handed (LH) components
of the physical quark arise as surface states on opposite
boundaries of a five dimensional slab with free boundary conditions in the
fifth direction. One five dimensional fermion  field is needed for every
physical quark. The surface fermions scheme has
also been discussed recently in \rf{mcih}.

  While Kaplan intended to propose a solution to the long-standing
problem of defining chiral gauge
theories on the lattice, the feasibility of reaching this goal is still
unclear~[16-19,22]. But the advantages of using chiral defect fermions
for lattice QCD are obvious. In particular, one can easily show that, in the
limit where the width of the five dimensional slab tends to infinity,
quark masses undergo only
{\it multiplicative} renormalization to all orders in perturbation
theory~\cite{bndr}.

  The reason for the absence of additive $O(1/a)$ corrections
to quark masses, is the vanishing of the overlap between the
RH and LH components of the quark's wave function.
At tree level, the tail of the RH wave function goes like
\beq
\label{tail}
  (1-Ma)^s\,.
\eeq
Here $s$ is the fifth coordinate which takes the values
$s=1,\ldots,2N$. The Dirac mass $M$ that appears in the five dimensional
fermion action obeys $0<Ma<1$. A similar expression applies to the LH
component, but with $s$ in \seeq{tail} replaced by $2N-s$.
Thus, the perturbative overlap of the LH and RH components vanishes
exponentially with increasing $N$.

  One can also introduce direct couplings between the LH and RH components
of every quark, by adding links that couple the layers $s=1$
and $s=2N$. These couplings are controlled by dimensionful
parameters $m_i$, where $i$ is a flavour index.
As shown in \rf{bndr}, the $m_i$ play the role of
multiplicatively renormalized quark masses.

  Taking advantage of the special properties of the model, one
can define axial currents whose divergences, for $m_i=0$, are completely
localized on the two middle layers $s=N$ and $s=N+1$.
As a result, the anomalous term in the Ward identities
of NSAS is governed by the small tail
of the quark's wave function at the center
of the five dimensional slab. On the other hand, the divergence of the
singlet axial current can couple to two gluons, giving rise in the limit
$N\to\infty$ to the expected axial anomaly~\cite{gjk,anml,nn2,mcih}.

  In this paper we extend the investigation of axial properties to include also
non-perturbative effects. Our main result is that  NSAS become exact in the
limit $N\to\infty$. To our knowledge, this is the first {\it non-perturbative
proof} of the restoration of NSAS.

  The restoration of NSAS occurs {\it before} the continuum limit is taken.
The limiting ``$N=\infty$'' formalism should be regarded as non-perturbative
regularization of QCD which is maximally symmetric under axial transformations.
Since the ``$N=\infty$'' formulation preserves NSAS while reproducing
correctly the singlet anomaly, it cannot be a strictly local regulator.
The non-locality arises from integrating out infinitely many heavy
four-dimensional fields, and we believe that it is mild
enough not to jeopardize the consistency of the continuum limit.

  This paper is organized as follows.
  In sect.~2 we give the definition of the model. Apart from the obvious gauge
and fermion fields, the model includes a set of massive five dimensional {\it
scalar} fields. These fields are necessary to cancel out some lattice artifacts
of the five dimensional fermions~\cite{sf,nn1}. Such scalar fields have often
been called Pauli-Villars (PV) fields in the literature, and we will continue
to use this terminology here. But it should be stressed that the scalar action
is non-negative, and so the partition function of the PV fields is
well-defined. The peculiar property of the PV fields is that, apart from a
different choice of boundary conditions, their action is the square of the
fermionic action.

  In sect.~3 we develop a transfer matrix formalism to represent the
fermionic partition function as well as correlation functions. The transfer
matrix technique was introduced in the context of chiral defect fermions
by Narayanan and Neuberger~\cite{nn2} and it relies on the work of
L\"uscher~\cite{l}. It proves particularly convenient for the investigation
of the model. Formulae are given both for finite $N$ and for the limiting
case $N\to\infty$.

  Sect.~4 contains a discussion  of the effective action $\seft(U)$ obtained by
integrating out the fermion and PV fields and taking the limit $N\to\infty$.
An interesting result is that, while being always real, $\exp\{-\seft(U)\}$ is
not necessarily positive. This behaviour can be explained on the basis of
familiar instanton results.

  The results of sect.~3 and~4 are valid for a fixed background gauge field,
provided the hamiltonian $\ch=-\log T$ where $T$ is the transfer matrix
has a unique ground state.  In sect.~5 we derive an analytical criterion
\seeq{detzm} which selects those gauge field configurations where $\ch$ has an
exact zero mode and, hence, the ground state is degenerate. Such backgrounds
allow for unsuppressed fermionic (and PV) propagation across in the fifth
dimension, and they contribute to the anomalous term in non-singlet axial Ward
identities.  The possibility of having light states {\it inside} the five
dimensional balk is related to the unconventional relative sign of the mass and
Wilson terms in the fermion action. (The hopping parameter is {\it
supercritical} ).  A simple example of a configuration which supports
anomalously light balk states is the {\it dynamical domain wall}.

  Sect.~6 contains the main results of this paper. Let
$Z(g_0,L,N)$ be the partition function on a finite five
dimensional lattice. The $N\to\infty$ limit of the partition function is
\beq
   Z^\infty(g_0,L) = \lim_{N\to\infty} Z(g_0,L,N) \,.
\label{limz}
\eeq
Both the four dimensional lattice $L^4$ and the bare coupling $g_0$ are kept
finite. We prove that
\beq
   Z^\infty(g_0,L) = \prod_{x,\m} \int dU_{x,\m}\, e^{-S_G(U)-\seft(U)} \,.
\eeq
$S_G(U)$ and $\seft(U)$ are defined in \seeqs{sg} and\seneq{seft} respectively.
In other words, one can interchange the order of integration over the group
variables and the  $N\to\infty$ limit. This result follows from the
$s$-independence of the gauge field, compactness
of the gauge field configuration space and \seeq{detzm}, which implies
that exact zero modes of $\ch$ exist only on a zero measure subspace.

  Repeating the same steps for non-singlet axial Ward identities, we prove that
the anomalous term in every Ward identity vanishes in the limit
$N\to\infty$, at fixed values of
$L$ and $g_0$. As expected, the same analysis gives rise to a non-zero
expression for the singlet anomaly.  We stress that NSAS are recovered {\it
before} the continuum limit $g_0\to 0$ is taken. Such a non-trivial result is
possible because the size of the unphysical fifth dimension $N$ is sent to
infinity. This is reflected in the operator expression for conserved
currents, which contains an infinite sum over the $s$-coordinate.

  A difficulty with the transfer matrix formalism, it that its efficient
implementation in numerical simulations would require the development of new
techniques. Instead, one may choose to put the fermions on a finite {\it five
dimensional} lattice. One should then choose an optimal value for
$N$ subject to the constraints dictated by computer performance.
To this end, it is important to have a realistic estimate of the
magnitude of anomalous effects on a finite five dimensional lattice.

  We believe that the bounds used in proving the existence of the chiral limit,
highly overestimate the true magnitude of anomalous terms. A detailed study of
the issue is beyond the scope of this paper.  However, we have decided to give
in Sect.~7 a short heuristic discussion, which is meant to give the reader some
feeling about the plausible magnitude of the anomalous term.  As we explain, we
believe that the magnitude of anomalous effects may turn out to be numerically
very small already for currently accessible five dimensional lattices.
Finally, some technical details are relegated to two appendices.

\vspace{5ex}
\noindent {\large\bf 2.~~Definition of the model}
\vspace{3ex}
\secteq{2}

  In this section we give the definition of lattice QCD with the surface
fermions variant of chiral defect fermions~\cite{bndr}.  We also define
physical quark operators as well as axial and vector currents~\cite{anml}
appropriate for the model. Most of the ingredients  have been introduced
previously, and we give them here to make the present exposition
self-contained.

   For definiteness, we take the physical case of four dimensions.
This means that the fermion and Pauli-Villars (PV) fields live on five
dimensional lattices, whereas the gluon fields are four dimensional. The
ordinary four coordinates, labeled $x_\m$, range from $1$ to $L$,
whereas the extra coordinate takes the values $s=1,\ldots,2N$
for the fermionic lattice. The PV lattice is only half as big, with
$s$ ranging from $1$ to $N$. The preferred boundary conditions in the four
ordinary dimensions are periodic or anti-periodic. Free boundary condition
in these directions would result in extra unwanted light states which can
propagate along the spatial boundaries.

  The above scheme is realized by requiring that the link
variables in the fermion and PV action obey $U_{x,s,5}=1$ and
$U_{x,s,\m}=U_{x,\m}$ independently of $s$.
The topology of the fifth dimension is taken to be a circle,
but the couplings which reside on the links
connecting the layers $s=2N$ and $s=1$ are proportional to a
parameter $~-m_i$ ($i=1,\ldots,N_f$ is a flavour index.
Also, we henceforth set the lattice spacing to $a=1$).
The case $m_i=1$ corresponds to antiperiodic boundary
conditions, where the model supports no light fermionic state.
The case $m_i=0$ corresponds to open boundaries, and it should
give rise to the physics of QCD with massless quarks by taking first the limit
$N\to \infty$ and then the continuum limit.

\newpage
\noindent  The partition function is
\bqry
\label{pf}
  Z  & = & Z(g_0,L,N,m_i) \NON
     & = & \prod_x
           \left( \prod_\m \int dU_{x,\m}
                  \prod_{s=1}^{2N} \int d\Bar\j_{x,s}\, d\j_{x,s}
                  \prod_{s'=1}^{N} \int d\f^\dagger_{x,s'}\, d\f_{x,s'}
           \right) e^{-S} \,.
\eqry
The action  is given by:
\beq
  S =  S_G(U) + S_F(\Bar\j , \j , U ) +
       S_{PV}(\f^\dagger, \f , U ) \,.
\eeq
Colour, flavour and Dirac indices will be suppressed throughout this paper
unless explicitly stated otherwise. We remind the reader that the PV fields
carry the same set of indices as the fermion fields. Also, we will usually
write $\cd U = \prod dU$ \etc as a shorthand for the corresponding measure.

  $S_G(U)$ is the pure gauge part of the action. The results of this paper
generalize to any compact Lie group with a non-negative lattice
action, which reduces to the appropriate gauge field action in the classical
continuum limit. To avoid irrelevant notational complications,
we will assume that the gauge group is $G=SU(N_c)$. In \seeq{pf}, $dU$ is
the normalized, invariant group measure,
and $S_G(U)$ is the the usual sum over plaquettes
\beq
   S_G = {1\over g_0^2} \sum_x \sum_{1\le\m<\n\le 4} \Re \tr(I-U_{x,\m\n})\,.
\label{sg}
\eeq
Here $U_{x,\m\n}$ is the plaquette variable.

  The fermion and PV actions contain a sum over all flavours.
The only difference between various flavours
can be in the mass parameter $m_i$. We give below the one flavour action.
The fermionic part has the following form
\beq
\label{Faction}
  S_F(\Bar\j , \j , U ) = -
  \sum_{x,y,s,s'} \Bar\j_{x,s} (D_F)_{x,s ; y,s'}\, \j_{y,s'} \,,
\eeq
where the fermionic matrix is defined by
\beq
\label{dirac}
  (D_F)_{x,s ; y,s'} = \d_{s,s'} D^\parallel_{x,y}  +
  \d_{x,y} D^\perp_{s,s'} \,,
\eeq
\bqry
\label{dpara}
  D^\parallel_{x,y} & = & \half \sum_\m
  \left(
  (1+\g_\m) U_{x,\m} \d_{x+\hat\m,y} +
  (1-\g_\m) U^\dagger_{y,\m} \d_{x-\hat\m,y}
  \right) \NON
  & & \quad + (M-4) \d_{x,y} \,,
\eqry
\beq
  D^\perp_{s,s'} = \left\{ \begin{array}{ll}
       P_R\, \d_{2,s'} - m P_L\, \d_{2N,s'} - \d_{1,s'} \,, &  s=1 \,, \\
       P_R\, \d_{s+1,s'} + P_L\, \d_{s-1,s'} - \d_{s,s'} \,, & 1<s<2N \,, \\
       - m P_R\, \d_{1,s'} + P_L\, \d_{2N-1,s'} - \d_{2N,s'} \,, & s=2N \,.
                   \end{array}\right.
\eeq
and
\beq
  P_{R,L} = \half(1\pm \g_5) \,.
\eeq

  Notice that $D^\perp_{s,s'}$ is independent of the gauge field.
Also, apart from the unconventional sign of the mass term,
$D^\parallel_{x,y}$ is the usual four dimensional gauge covariant Dirac
operator for massive Wilson fermions.

  When $m_i=0$ the spectrum of surface states contains one
RH Weyl fermion near the boundary $s=1$ and one LH Weyl fermion near the other
boundary for every five dimensional fermion field.
These Weyl fermions have the same coupling to the
gauge field, and so they  in fact describe $N_f$ quarks. If we ignore the
exponentially small overlap between the tails of the LH and RH surface states,
then these states describe {\it massless} quarks.   Switching $m_i$ on, we now
mix the RH and  LH components of each quark  and provide it with a Dirac mass
$\Bar{m}_i$ which is proportional to $m_i$. At tree level one has~\cite{bndr}
\beq
\label{mm}
  \Bar{m}_i = M(2-M)m_i \,.
\eeq

  The PV fields~\cite{sf,nn1} are needed to cancel the contribution of heavy
fermion modes to the effective action $\seff(U)$.
This contribution, while formally being local in the
continuum limit, is proportional to $N$. (Every five dimensional fermion field
describes one light quark field and $2N-1$ four dimensional fields whose mass
is
of the order of the cutoff). If one does not subtract the contribution of the
massive fields by hand, that lattice artifact will dominate the effective
action in the limit $N\to\infty$.

   Let us denote the dependence of the Dirac operator in \seeq{dirac} on $m_i$
and on the number of sites in the $s$-direction by $D_F=D_F(2N,m_i)$. The
PV fields live on a five dimensional lattice
with $N$ sites in the $s$-direction, and using the above
notation, the PV action is
\beq
  S_{PV}(\phi^\dagger, \phi , U )=
  \sum_{x,y,z,s,s',s''} \phi^\dagger_{x,s}\,
  D^\dagger_F(N,1)_{x,s;z,s''}\,
   D_F(N,1)_{z,s'';y,s'}\, \phi_{y,s'}\,.
\label{PVaction}
\eeq
The second order operator in \seeq{PVaction} is the square of the Dirac
operator
on a smaller lattice. The choice $m_i=1$ for the PV fields prevents the
appearance of light scalar modes on the layers $s=1$ and $s=N$.
As will be shown below, this  choice of $S_{PV}$ ensures that $\exp\{-\seff\}$
remains finite in the limit $N\to\infty$.

 Let us denote by $\car$ the reflection relative
to the hyper-plane $s=N+1/2$
\beq
\car\, \j_{x,s} =  \j_{x,2N+1-s} \,.
\eeq
The Dirac operator $D_F$ of \seeq{dirac} satisfies the following
identity
\beq
\label{reflection}
 \g_5 \car\, D_F\, \g_5 \car = D^\dagger_F \,.
\eeq
This identity is a generalization of a similar relation obeyed by the
four dimensional Dirac operator for Wilson fermion, which reads
\beq
  \g_5\, D^\parallel\, \g_5 = (D^\parallel)^\dagger \,.
\eeq
As in the case of Wilson fermions, \seeq{reflection} plays an important role
in establishing the positivity of  pion correlators (see App.~B).

  \seEq{reflection} implies that the operator $\g_5 \car D_F$ is
hermitian. One has $\det(\g_5 \car)=1$ trivially, and so
\beq
   \det(D_F)=\det(\g_5 \car D_F) \,.
\eeq
As a result, the fermionic determinant is real. However, one cannot
conclude that the fermionic determinant is necessarily positive. The fermionic
determinant can in fact change sign, due to level crossing of an odd number of
states (see below).

  We comment in passing that one can use the hermitian operator $\g_5 \car D_F$
in the definition of the fermionic action instead of $D_F$. This is facilitated
by the unitary change of variables (recall that $\j$ and $\Bar\j$ are
independent variables in euclidean space)
\bqry
   \j & \to & \j'=\j \,, \NON
   \Bar\j & \to & \Bar\j'=\Bar\j\g_5\car \,.
\label{j'}
\eeqr
A similar definition (involving $\g_5$ only) has been found useful in the
construction of the euclidean partition function of continuum supersymmetric
QCD~\cite{susy}. Notice that the uneven
treatment of $\j$ and $\Bar\j$ in \seeq{j'} gives rise to unconventional
expressions for the axial and vector currents. Since we will not use this
formulation any further we do not give the details here.

  In the rest of this paper we assume $m_i=m$, $i=1,\ldots,N_f$.
The five dimensional fermion action is invariant under global
$U(N_f)$ symmetry. The conserved {\it five dimensional} current has the
following components. For $\m=1,\ldots,4$
\bqry
  j^a_\m(x,s) & = & \half\left(
        \Bar\j_{x,s} (1+\g_\m) U_{x,\m} \l^a
         \j_{x+\hat\m,s}\right.  \NON
    & & \left. \quad\quad
       - \Bar\j_{x+\hat\m,s} (1-\g_\m) U^\dagger_{x,\m} \l^a \j_{x,s}
                  \right)\,, \quad 1\le s \le 2N \,.
\label{5dcrnt}
\eqry
As for the fifth component, we define
\beq
  j^a_5(x,s) = \left\{ \begin{array}{ll}
    \Bar\j_{x,s} P_R \l^a \j_{x,s+1}  - \Bar\j_{x,s+1} P_L \l^a \j_{x,s}\,,
    & 1\le s < 2N\,, \\
    \bar\j_{x,2N} P_R \l^a \j_{x,1} - \bar\j_{x,1} P_L \l^a \j_{x,2N}\,,
    & s=2N \,.
     \end{array}\right.
\eeq
This five dimensional current satisfies the continuity equation
\beq
  \sum_\m\D_\m\, j^a _\m(x,s) =
     \left\{ \begin{array}{ll}
          - j^a_5(x,1) - m j^a_5(x,2N)\,,      &  s=1 \,,     \\
          - \D_5\, j^a_5(x,s)\,, &  1<s<2N \,, \\
            j^a_5(x,2N-1) + m j^a_5(x,2N)\,,      &  s=2N \,.
     \end{array}\right.
\label{5cons}
\eeq
Here
\beq
  \D_\m f(x,s) = f(x,s)-f(x-\hat\m,s)\,,
\eeq
\beq
  \D_5 f(x,s) = f(x,s)-f(x,s-1)\,.
\eeq
$\l^a$ is a flavour symmetry generator. Notice the special form of the boundary
terms in the continuity equation.

  We now give the definitions of {\it four dimensional} vector and axial
currents~\cite{anml}. There is a unique set of conserved vector currents,
given by
\beq
  V^a_\m(x)=\sum_{s=1}^{2N} j^a _\m(x,s) \,.
\label{4vec}
\eeq
Conservation of the vector current $V^a_\m$ follows from \seeq{5cons}.

  On the other hand, there is a lot of arbitrariness in defining axial
transformations in the model. Any transformation which assigns opposite
charges to the LH and RH chiral modes will reduce to the familiar axial
transformation in the continuum limit.

  Here we take advantage of the global separation of the LH and RH modes
in the $s$-direction. We define our axial transformation to act {\it
vectorially} on a given four dimensional layer, but we assign opposite charges
to fermions in the two half-spaces
\bqry
\label{axial}
  \d^a_{A} \j_{x,s}     & = & + i q(s) \l^a \j_{x,s} \,, \\
  \d^a_{A} \Bar\j_{x,s} & = & - i q(s) \Bar\j_{x,s} \l^a \,,
\eqry
where
\beq
  q(s) = \left\{ \begin{array}{rr}
              1\,,  &  1\le s \le N \,, \\
              -1\,,  &  N <s \le 2N   \,.
           \end{array}\right.
\eeq
The corresponding axial currents are
\beq
\label{axcur}
  A^a_\m(x) = - \sum_{s=1} ^{2N} \sign(N-s + \half)\, j^a_\m (x,s) \,.
\eeq

  For $m=0$, the non-invariance of the action under the
transformation\seneq{axial} resides entirely in the coupling between
the  layers $s=N$ and
$s=N+1$. For $m\ne 0$, there is an additional contribution coming from the
direct coupling between the layers $s=1$ and $s=2N$. As a result, the axial
currents satisfy the following divergence equations
\beq
\label{div}
   \D_\m A^a_\m(x) = 2m J^a_5(x) + 2 J^a_{5q}(x) \,,
\eeq
where
\beq
\label{psc}
   J^a_5(x) = j^a_5(x,2N) \,,
\eeq
\beq
   J^a_{5q}(x) = j^a_5(x,N) \,.
\eeq

  In order to understand the physical content of \seeq{div} let us define
quark operators as follows
\bqry
   q_x & = & P_R\, \j_{x,1} + P_L\, \j_{x,2N} \NON
   \Bar{q}_x & = & \bar\j_{x,2N} P_R  + \bar\j_{x,1} P_L  \,.
\label{quark}
\eqry
These operators have a finite overlap with the surface states in the chiral
limit $N\to\infty$, and so they can play the role of bare quark fields. There
are of course many other operators which are localized near the boundaries and,
hence, can interpolate quark states. Our choice \seeq{quark} is the simplest
possible one, and it leads to considerable simplification of the expressions
for correlation functions.

  In terms of the quark fields, $J^a_5$ takes the familiar form
\beq
   J^a_5(x) = \bar{q}_x \g_5 \l^a q_x \,.
\eeq
Thus, up to the finite normalization factor in \seeq{mm},
the $J^a_5$ term on the \rhs of \seeq{div} is the expected
contribution from a classical mass term. $J^a_{5q}$ represents an additional,
quantum breaking term. Below we will be interested in axial Ward
identities of the general form
\bqry
\label{wi}
   \D_\m \vev{A^a_\m(x)\, O(y_1,y_2,\ldots) } & = &
     2m \vev{J^a_5(x)\, O(y_1,y_2,\ldots) }  \NON
   & &  + 2 \vev{J^a_{5q}(x)\, O(y_1,y_2,\ldots)} \NON
   & &  + i \vev{\d^a_A\, O(y_1,y_2,\ldots)}\,,
\eqry
where $A^a_\m$ is a non-singlet axial current. Our aim is to show that, if
$O(y_1,y_2,\ldots)$ involves only the quark operators of \seeq{quark} then the
anomalous term
\beq
   \vev{J^a_{5q}(x)\, O(y_1,y_2,\ldots)}
\label{bad}
\eeq
vanishes in the limit $N\to\infty$. An especially important case is when the
operator $O$ is itself a pseudo-scalar density $O=J^b_5(y)$. The
corresponding Ward identity
\bqry
\label{pionms}
   \D_\m \vev{A^a_\m(x)\, J^b_5(y) } & = &
     2m \vev{J^a_5(x)\, J^b_5(y) }  \NON
   & &  + 2 \vev{J^a_{5q}(x)\, J^b_5(y) } \NON
   & &  - \d_{x,y} \vev{ \bar{q}_y \{\l^a,\l^b\} q_y }\,,
\eqry
governs the pion mass.

\newpage
\noindent {\large\bf 3.~~Transfer matrix formalism}
\vspace{3ex}
\secteq{3}

  It is convenient to discuss non-perturbative effects using the transfer
matrix formalism. This technique was adapted to domain wall fermions in
\rf{nn2}. Here we in fact have a constant five dimensional mass $M$, and the
only deviation compared to \rf{l} is in the $m_i$-dependent value of the
couplings on the links connecting the layers $s=1$ and $s=2N$.
In this section and the next one we will assume that the background gauge
field $U_{x,\m}$ is fixed.

  A simple generalization of the result of \rf{l} gives rise to the following
second quantized expression for the Grassmann path integral
\bqry
\label{fdet}
 \det D_F(2N,m)  & = & \int \cd\Bar\j\cd\j\, e^{-S_F(\Bar\j,\j,U)} \NON
                 & = & (\det B)^{2N}\, \tr T^{2N} \co(m) \,.
\eqry

  The second quantized transfer matrix $T$ acts in a Fock space spanned by the
action of fermionic creation operators $\hat{a}^\dagger_x$ on a vacuum state
$\sket{0}$ annihilated by $\hat{a}_x$.  The operators $\hat{a}_x$ and
$\hat{a}^\dagger_x$ satisfy canonical anticommutation relations. They live on
the sites $x_\m$ of a four dimensional lattice of size $L^4$, and they also
carry Dirac, colour and flavour indices which we have suppressed. The
decomposition of $\hat{a}_x$ into RH and LH components is
\beq
  \hat{a}= \left(\begin{array}{c}
             \hat{c} \\
             \hat{d}^\dagger
           \end{array}\right)
\eeq
The transfer matrix is defined by
\beq
   T = e^{-\hat{a}^\dagger H \hat{a} } \,,
\eeq
where
\beq
  e^{-H} = \left( \begin{array}{rl}
              B^{-1} & B^{-1}C \\
              C^+ B^{-1} & C^+ B^{-1} C +B
        \end{array} \right)
\label{exph}
\eeq
\beq
   B_{x,y} = (5-M)\d_{x,y} -\half \sum_\m \left[\d_{x+\hat\m,y} U_{x,\m}
             + \d_{x-\hat\m,y} U^\dagger_{y,\m} \right] \,,
\eeq
\beq
   C_{x,y} = \half \sum_\m \left[\d_{x+\hat\m,y} U_{x,\m}
             - \d_{x-\hat\m,y} U^\dagger_{y,\m} \right] \s_\m \,,
\eeq
and $\s_\m=(i,\vec\s)$. An important identity is
\beq
  e^{-H}=K K\dg \,,
\label{kkdg}
\eeq
where
\beq
   K = \left( \begin{array}{cc}
              B^{-1/2}      &  0       \\
              C\dg B^{-1/2} &  B^{1/2}
       \end{array}\right) \,.
\label{k}
\eeq
The last equation implies that the matrix operator $H$ is well-defined.
Notice also that $D^\parallel$ can be
expressed in terms of $B$ and $C$ as follows
\beq
\label{dpr}
   D^\parallel = \left(\begin{array}{cc}
                        1-B    & C     \\
                        -C\dg  & 1-B
                 \end{array}\right)\,.
\eeq
The operator $\co(m)$ contains all the $m$-dependence, and it is given by
\beq
\label{com}
  \co(m) = \prod_n (\hat{c}_n \hat{c}^\dagger_n + m\hat{c}^\dagger_n \hat{c}_n)
             (\hat{d}_n \hat{d}^\dagger_n + m \hat{d}^\dagger_n \hat{d}_n) \,.
\eeq
In this equation, $n$ is a generic name for all indices.  The special cases
$m=0$ and $m=1$ deserve special attention. For $m=1$, $\co(m)$ becomes the
identity operator, whereas for $m=0$ it is a projection operator on a different
ground state $\sket{0'}$ annihilated by all the $\hat{c}$-s and $\hat{d}$-s.
The relation between $\sket{0}$ and $\sket{0'}$ is
\beq
\label{00'}
   \ket{0} =\prod_n \hat{d}^\dagger_n \ket{0'} \,.
\eeq

  Both $\sket{0}$ and $\sket{0'}$ are ``kinematical'' ground states which can
serve as convenient reference states in the construction of the Fock space.
But none of them play a special role dynamically. Since we use
creation and annihilation operators in the coordinate basis,  both
$\sket{0}$ and $\sket{0'}$ are very different from the filled Dirac sea even
in the case of free fermions.

  Like $D^\parallel$, the hermitian operator $H$ is a $N_t\times N_t$ matrix,
where $N_t= 4 N_c N_f L^4$.
Let $R$ be the unitary matrix which  diagonalizes $H$
\beq
   \sum_l H_{ml} R_{ln}  = E_n R_{mn} \,.
\eeq
Corresponding to the splitting  into two-by-two matrices in Dirac space in
\seeq{exph}, we write $R$ as two $\half N_t\times N_t$ matrices $P$ and $Q$.
We will assume that the columns of $R$ are ordered according to their
eigenvalues, with the negative eigenvalues on the left. Accordingly, we further
decomposed $P$ and $Q$ into submatrices which contain the positive and negative
eigenvectors
\beq
   R  = \left( \begin{array}{cc}
                P^- & P^+ \\
                Q^- & Q^+
          \end{array} \right)\,.
\eeq
$P^\mp$ and $Q^\mp$ are $\half N_t\times N^\mp$ matrices, where
$N^+ + N^- = N_t$.

  The ground state of of the second quantized operator
$\ch=\hat{a}^\dagger H \hat{a}$ is obtained by filling all negative energy
states
\beq
\label{gs}
  \ket{0_H}= \prod_{i=1}^{N^-} \left(
                \hat{c}^\dagger_{l_i} P^-_{l_i,i}  +
                \hat{d}_{l_i}         Q^-_{l_i,i}
             \right) \ket{0} \,.
\eeq
In the limit $N \to\infty$, $T^{2N}$ is proportional to a projector on the
ground state of $\ch$
\beq
\label{infN}
   T ^{2N} \to \ket{0_H} \l_{\rm max}^{2N} \bra{0_H} ~,\quad \quad
N\to\infty\,,
\eeq
where
\beq
\l_{\rm max} = \exp\left\{-\sum_{i=1}^{N^-} E_i \right\}\,.
\eeq

  We now turn to the scalar PV fields. The action is bilinear in these
fields, and so
\bqry
  \int\cd\f\cd\f^\dagger\, e^{-S_{PV}(\f^\dagger,\f,U)}
   & = & \det^{-1} \left( D\dg_F(N,1) D_F(N,1) \right) \NON
   & = & (\det B)^{-2N} \left(\tr T^{N} \right)^{-2}\,.
\label{pvdet}
\eqry
In going from the first to the second row we have used \seeq{fdet} and
substituted $m=1$.

  The effective action $\seff(U)$ is defined by integrating out both fermion
and PV fields. Using \seeqs{fdet} and\seneq{pvdet} we find
\beq
\label{seff}
  \exp\{-\seff\} = { \tr T^{2N} \co(m) \over \left(\tr T^{N} \right)^2 } \,.
\eeq
In the limit $N \to\infty$  the effective action becomes
\bqry
\label{seft}
  \exp\{-\seft\} & \equiv &  \lim_{N\to\infty} \exp\{-\seff\} \NON
                 &   =    &  \bra{0_H} \co(m) \ket{0_H} \,.
\eqry

  \seEq{seft} is valid for a fixed background field provided the ground state
of $\ch$ is non-degenerate. The limiting expression \seeq{seft}
is completely well defined here, and it is free of any subtleties of
the kind encountered in trying to define chiral gauge theories on the lattice
using chiral defect fermions~\cite{anml}.

\vspace{5ex}
\noindent {\large\bf 4.~~The effective action}
\vspace{3ex}
\secteq{4}

  As a first step, we want to study the behaviour of
$\sbra{0_H}\co(m)\sket{0_H}$
under various conditions. The physically interesting case is $m\ll 1$.
In the case $m=1$ one has $\sbra{0_H} \co(1) \sket{0_H}=1$. The reason for this
trivial result is the subtraction of the balk effect through the PV fields.
By contrast, in the opposite limit $m=0$ one has
\beq
\label{ovrlp}
  \exp\{-\seft\} = \abs{\Braket{0_H}{0'}}^2\,,\quad\quad m=0 \,.
\eeq
Thus, in a model of massless chiral defect fermions, the physical information
is not in the trace of the transfer matrix but, rather, in the overlap of its
ground state with some other state. This was first found by Narayanan
and Neuberger for domain wall fermions, where the overlap formula
reads~\cite{nn2}
\beq
\label{ovrlpdm}
  \exp\{-\seft(\mbox{domain wall})\} = \abs{\Braket{0_{H_+}}{0_{H_-}}}^2 \,.
\eeq
Here $H_\pm$ are the hamiltonians corresponding to the two sides of the domain
wall. Hence, one has to compare two {\it dynamical} ground states. But all the
non-trivial dynamics of the model is contained in the $H_+$ hamiltonian. In
the continuum limit, \seeqs{ovrlp} and\seneq{ovrlpdm} should describe the same
physics. It is therefore advantageous to work with the surface fermion scheme,
where one has to calculate the overlap of $\sket{0_H}$ with a fixed reference
state $\sket{0'}$.

  An explicit expression for the $m=0$ overlap \seeq{ovrlp} can be easily
written down~\cite{nn2}. We first comment that for free fermions, as well as
for perturbative gauge field configurations, the numbers of positive and
negative eigenvalues of $H$ are equal $N^\mp = \half N_t$. In this case
$P^\mp$ and $Q^\mp$ are square matrices. Using \seeq{00'} it follows that
only the $\hat{d}$-dependent terms in \seeq{gs} contribute to the  overlap
\seeq{ovrlp}. Taking into account the anticommuting character of the fermionic
operators one arrives at
\beq
   \Braket{0_H}{0'} = \det Q^- \,.
\eeq
The phase ambiguity in defining the columns of $Q^-$ is irrelevant, because
only the modulus of $\det Q^-$ enters \seeq{ovrlp}.

  For non-perturbative configurations there may be level crossing, resulting
in $N^\mp\ne\half N_t$. In this case the $m=0$ overlap
vanishes identically. As discussed in \rf{nn2} this is
a welcomed phenomenon, which signals that the chiral defect fermion model
can reproduce instanton effects.

  We now want to generalize the explicit expression for
$\sbra{0_H}\co(m)\sket{0_H}$
to $m\ne 0$. We first notice that $\co(m)$ can be expanded as
\beq
\label{oexp}
 \co(m)= \sum_{k=0}^{N_t} m^k \sum_{l+n=k} {1\over l!\, n!}
            \hat{c}^\dagger_{i_1} \ldots \hat{c}^\dagger_{i_l}
            \hat{d}^\dagger_{j_1} \ldots \hat{d}^\dagger_{j_n}
            \ket{0'}\bra{0'}
            \hat{c}_{i_1} \ldots \hat{c}_{i_l}
            \hat{d}_{j_1} \ldots \hat{d}_{j_n} \,.
\eeq
In \seeq{oexp}, $m^k$ multiplies a sum over orthogonal projection operators
whose number grows like $N_t^k/k!$. Recall that $N_t$ is proportional to the
four-volume $V=L^4$.  In calculating
the $m$-expansion of the effective action, at the $k$-th order we will
therefore encounter $O(V^k/k!)$ terms,
where the magnitude of each term is bounded by $m^k$.
This is in agreement with the anticipated behaviour of a system which
undergoes spontaneous symmetry breaking. The finite volume partition function
is analytic in $m$, but in the infinite volume limit singularities may
appear because the product $mV$ diverges.

  Returning to the calculation of $\sbra{0_H}\co(m)\sket{0_H}$, let us denote
\beq
  \D= \left| \half N_t - N^- \right| \,.
\eeq
The first term in the expansion \seeq{oexp} which contributes to
$\sbra{0_H}\co(m)\sket{0_H}$
is the term with $k=\D$. This immediately implies that
\beq
  \bra{0_H} \co(m) \ket{0_H} = O(m^{\D})\,.
\eeq

  Moreover, for sufficiently small $m$, the {\it sign} of
$\sbra{0_H}\co(m)\sket{0_H}$ is determined by the sign of $m^{\D}$. Hence,
$\sbra{0_H}\co(m)\sket{0_H}$ will be negative for $m<0$ and odd $\D$. For
example, in the one flavour case $\sbra{0_H}\co(m)\sket{0_H}$ is negative
for $m<0$ if an odd number of level crossing took place.

  This behaviour is not unexpected and, in fact, it is in agreement with the
familiar instanton result~\cite{thooft}. In the one flavour case, the
fermionic determinant in an instanton background is proportional to $m$,
and so it changes sign if $m$ does. The gauge field's effective measure
$\exp\{-S_G-\seft\}$ is therefore real, but not always positive.
We still expect the partition function to be strictly positive
when one approaches the continuum limit, because configurations with a non-zero
topological charge are rare. But we are unable to prove the positivity of
the partition function in a completely general way. Special cases where
$\exp\{-\seft\}$ is strictly positive include $m>0$,
or $m\ne 0$ and even $N_f$.

  An explicit expression for $\sbra{0_H}\co(m)\sket{0_H}$ is more easily
obtained using the definition \seeq{com}. Straightforward application of the
canonical anticommutation relations gives rise to
\beq
\label{omgs}
   \co(m)\ket{0_H} = m^{\half N_t} \prod_{i=1}^{N^-} \left(
               m \hat{c}^\dagger_{l_i} P^-_{l_i,i}  +
               m^{-1} \hat{d}_{l_i}    Q^-_{l_i,i}
             \right) \ket{0} \,.
\eeq
Introducing the $N_t\times N^-$ matrix
\beq
  R^-(m) = m^{\half N_t} \left( \begin{array}{c}
                              m P^-  \\
                              m^{-1} Q^-
                    \end{array} \right)\,,
\eeq
we find
\beq
  \bra{0_H}\co(m)\ket{0_H} = \det R^{\dagger -}(1)\, R^-(m) \,.
\eeq
In the case $m>0$ one can use the relation $\co(m) = \co(m^\half)\co(m^\half)$
to write
\beq
  \bra{0_H}\co(m)\ket{0_H} = \det R^{\dagger -}(m^\half)\,
                                  R^-(m^\half) \,,
\eeq
which is manifestly positive.

  The transfer matrix formalism can also be used to write down expressions
for correlation functions. The correlation functions for the quark
operators\seneq{quark} take a particularly simple form. The rules for
their construction, as well as some explicit examples, are given in Appendix~A.
The transfer matrix formula for axial Ward identities is discussed
also in Sect.~6.

\vspace{5ex}
\noindent {\large\bf 5.~~The dynamical domain wall}
\vspace{3ex}
\secteq{5}

    In the previous section we discussed some properties of the effective
action under the assumption that the ground state of $\ch$ is
non-degenerate. The ground state will be degenerate if $\ch$ has an exact
zero mode or, equivalently, if the matrix
$e^{-H}$ has a unit eigenvalue
\beq
\label{psi0}
e^{-H} \psi_0 =\psi_0 \,.
\eeq

  If \seeq{psi0} holds for a particular background field we expect strong
correlations accross the five dimensional slab. In particular, the
anomalous term in NSAS Ward identities receives its dominant contribution
from such configurations. It is therefore important to identify
these configurations and to understand their properties.

  The matrix operator $e^{-H}$ is non-local, and to understand better
the physical content of \seeq{psi0} we want to find a simpler
equation satisfied by the zero mode $\psi_0$.
Using \seeqs{kkdg}, (\ref{k}) and\seneq{dpr} it is straightforward to show that
\bqry
\label{zm}
   0 & = & (K\dg  - K^{-1})\psi_0 \NON
     & = & B^{-1/2} \g_5 D^\parallel \psi_0 \,.
\eqry
Since $B$ is a positive definite operator, we conclude that $\psi_0$ is a zero
mode of the hermitian operator $\g_5 D^\parallel$. Therefore, a necessary and
sufficient condition for the existence of a zero mode is
\beq
  \det(\g_5 D^\parallel) = 0 \,.
\label{detzm}
\eeq
Notice that the \lhs of \seeq{detzm} is a polynomial in the group variables
$U_{x,\m}$.

  Since $D^\parallel$ is a massive lattice Dirac operator, one may ask
whether \seeq{psi0} has any non-trivial solutions at all. An indirect way to
argue that such solutions should exist, is to observe that this is a necessary
condition for $\sbra{0_H}\co(m)\sket{0_H}$ to reproduce the instanton results.
The vanishing of this expectation value for $m=0$ requires that level crossing
should occur in the spectrum of $H$. At the crossing point one has a solution
of \seeq{psi0}. This observation was made by Narayanan and
Neuberger~\cite{nn2},
who also found numerically solutions of \seeq{psi0}.

  The reason why solutions to \seeq{psi0} exist in spite of the mass term
present in $D^\parallel=D^\parallel(M)$,
is the unconventional sign of that mass term.
For comparison, the conventional Dirac operator for a massive Wilson fermion
is $D^\parallel (-M)$ in our notation.  The ``wrong'' sign of the mass term
relative to the Wilson term, implies that the sum of the two terms is not
a positive definite operator. As a result, Vafa-Witten bounds~\cite{vw}
cannot be established here. On the other hand,
if one were to choose the conventional relative
sign, then a Vafa-Witten bound could be establish for
the propagation of fermions in {\it all} directions. This, in turn, would
imply the absence of any light states in the model. Indeed, one can easily
check that the massless surface modes disappear for $M<0$. We comment that
in the domain wall case, too, all the non-trivial dynamics occurs
on that side of the wall where one has the ``wrong'' relative sign.

  In the absence of gauge fields, as well as in perturbation theory, the only
zero modes of the five dimensional Dirac operator $D_F$ are the ones
discussed in refs.~\cite{k,jsc,bndr}. But with dynamical gauge
fields, other zero modes can exist.  Since the gauge field is
$s$-independent, any zero mode of the four dimensional operator $D^\parallel$
will also be a zero mode of the five dimensional operator
$D_F$.  This is true up to boundary effects, which can be ignored here because
the new type of zero modes in not localized in the $s$-direction.

  An interesting example is provided by the {\it dynamical domain wall}.
For simplicity we consider here a $U(1)$ gauge group. (The same configuration
exists also for $SU(2)$. For general $SU(N)$ one
can simply embed the below configuration in some $SU(2)$ subgroup).
We consider the following configuration of link variables. We let
$U_{x,\m}=1$ for $\m=1,3,4$. For $U_{x,2}$ we take
\beq
\label{ddm}
   U_{x,2} = \left\{ \begin{array}{rr}
                        1\,,  &  x_1 < x_1^0 \,, \\
                       -1\,,  &  x_1 \ge x_1^0 \,.
           \end{array}\right.
\label{conf}
\eeq
Notice that this is essentially a two dimensional configuration. \seEq{conf}
describes a wall of magnetic flux, with one unit of flux going through each
plaquette to the left of the hyperplane $x_1=x_1^0$.

  We now consider the ansatz
\beq
\label{j0}
   \j_0=\half(1-\g_1)\J_0(x_1) \,.
\eeq
One can easily check that the following is a zero mode of $D^\parallel$
\beq
    \J_0 (x_1) = \left\{ \begin{array}{lr}
                        (1-M)^{x_1^0-x_1}\,,  &  x_1 < x_1^0 \,, \\
                        (3-M)^{x_1^0-x_1}\,,  &  x_1 \ge x_1^0 \,.
                 \end{array}\right.
\eeq

  This is the simplest case of a dynamically generated zero mode. Other gauge
field configurations with a topological character may also support zero modes
of $D^\parallel$. These include for example lattice analogs of the static
string-like singularity discussed in \rf{ch}. The zero modes observed in
\rf{nn2} are actually of that type.

  Finally, we recall the close relationship between the fermionic and PV
action.  Whenever the hermitian operator $\g_5 D^\parallel$ has light states,
the same will be true for the PV fields. Therefore, anomalously strong
correlations in the $s$-direction arise simultaneously for the fermions and
the PV fields.  However, the two contributions
to the anomalous term {\it do not} cancel each other in general.

  In the Ward identity \seeq{pionms} which
governs the pion mass, the anomalous term is the correlator of two pseudoscalar
densities. As a generalization of similar results for Wilson fermions, one can
prove in the present model  the positivity of a whole family of correlators of
pseudoscalar densities. The details can be found in App.~B. As a result, the
anomalous term in \seeq{pionms} is strictly positive for finite values of $N$
and $g_0$. The vanishing of the
anomalous term in the limit $N\to\infty$ can only arise from the vanishing
of its absolute value on almost the entire gauge field configuration space.

\vspace{5ex}
\noindent {\large\bf 6.~~The chiral limit}
\vspace{3ex}
\secteq{6}

  The characterization\seneq{detzm} of gauge field configurations which support
exact fermionic zero modes, allows us to prove the following important
result (see \seeq{limz})
\beq
  \lim_{N\to\infty} Z(g_0,L,N,m) = \int \cd U\, e^{-S_G(U)}\,
                                 \bra{0_H} \co(m) \ket{0_H} \,.
\label{limzz}
\eeq
In technical terms, \seeq{limzz} means that we can interchange
the order of the $N\to\infty$ limit and the integration over the group
variables.  As we discuss below, the physical content of \seeq{limzz} is
that correlation functions obey clustering in the $s$-direction.

  Let us first introduce some terminology. The gauge field configuration space
is $\cg=(SU(N_c))^{4L^4}$. An element of $\cg$, \ie a particular
configuration of link variables will be denoted by $\cu=\{U_{x,\m}\}$.
We let $\cg_0\subset\cg$ be the subspace of all configurations which satisfy
condition\seneq{detzm}.

  In this section we make the  technical assumption $0\le m\le 1$. (The reader
should not confuse this condition with the similarly looking one $0<M<1$, which
we enforce for entirely different reasons). The condition $m\ge 0$ ensures the
positivity of $\sbra{0_H} \co(m) \sket{0_H}$. In the continuum limit we expect
to recover the symmetry under $m\to -m$, and so choosing a non-negative $m$
should not lead to any restrictions on the physical content of the model. The
condition $|m|\le 1$  implies $\parallel\!\co(m)\!\parallel\,\le 1$. This gives
rise to less cumbersome expressions for some of the bounds below. As we have
already explained, the physically interesting case is $|m| \ll 1$.

  The proof of \seeq{limzz} is simple, and it relies on the following
ingredients. (a) The gauge field is $s$-independent, and the configuration
space $\cg$ is compact. (b) According to standard results in calculus, the
subspace $\cg_0$ has a zero measure.

  Consider an element $\cu\in\cg-\cg_0$, and let
$E_n$ be the eigenvalues of $H(\cu)$. We define
\beq
  E_0(\cu) = \min\{ \, |E_n| \, \} \,.
\eeq
Since $H(\cu)$ is a finite dimensional matrix, $E_0(\cu)$ is well-defined, and
since $\cu\in\cg-\cg_0$, one has $E_0(\cu)>0$.

  We use this information to put a bound on the difference between the \rhs
of \seeq{seff} and the \rhs of \seeq{seft}. Separating the ground
state contribution from the rest we find
\beqrabc{6.3}
 & &  \left| \tr T^{2N} \co(m) / \left(\tr T^{N} \right)^2 -
      \bra{0_H} \co(m) \ket{0_H} \right| \le \label{inqa} \\
 & \le & \l_{\rm max}^{-2N}\, \tr'\, T^{2N} +
      \left| \l_{\rm max}^{2N}\, (\tr T^N )^{-2} - 1 \right| \label{inqb} \\
 & \le & 2^{N_t} \left[ e^{-2NE_0(\cu)}
       + e^{-NE_0(\cu)} (2+ 2^{N_t} e^{-NE_0(\cu)}) \right] \,. \label{inqc}
\eeqr
\setcounter{equation}{3}
\renewcommand{\theequation}{6.\arabic{equation}}
  The notation $\tr'$ means that the ground state contribution is excluded.
Notice that $2^{N_t}$ is the dimensionality of the fermionic Fock space.
The last row of the above inequality highly overestimates its first row.
Nevertheless, it will be sufficient for our purpose, because it implies that
expression\seneq{inqa} vanishes in the limit $N\to\infty$ for $E_0(\cu)>0$.

  We now have to show that for arbitrary $\e>0$, there exist $N_\e$ such that
\beq
  \left| Z(g_0,L,N,m) - \int \cd U\, e^{-S_G(U)}\,
         \bra{0_H} \co(m) \ket{0_H} \right| \le \e \,,
\label{epsln}
\eeq
for every $N\ge N_\e$. To this end, we divide the group integration into
two parts
\beq
   \int_\cg \cd U = \int_{\cg_\e} \cd U + \int_{\cg-\cg_\e} \cd U \,.
\eeq
$\cg_\e$ is an open covering of $\cg_0$ whose volume is less than $\e/4$.
Such an open covering can always be found. Furthermore, the \rhs of both
\seeqs{seff} and\seneq{seft} is always bounded by one. Hence, the contribution
to the \lhs of inequality\seneq{epsln} coming from the integration over
$\cg_\e$ is bounded by $\e/2$.

  Now let us denote
\beq
   \bar{E}_0 = \min\{ E_0(\cu)\: | \: \cu\in \cg-\cg_\e \} \,.
\eeq
Since $\cg-\cg_\e$ is compact, the minimum exists and satisfies
$\bar{E}_0>0$. It now follows that the contribution to the \lhs
of inequality\seneq{epsln} coming from the integration over $\cg-\cg_\e$ is
bounded by expression\seneq{inqc} with $E_0(\cu)$ replaced by $\bar{E}_0$.
The existence  of $N_\e$ such
that inequality\seneq{epsln} holds now follows from that fact that
$\bar{E}_0>0$. This completes the proof of \seeq{limzz}.

  Similar results can be established for correlation functions.  In order to
prove the restoration of NSAS in the limit $N\to\infty$, we have to show that
the anomalous term\seneq{bad} in the Ward identity \seeq{wi} vanishes for any
operator $O(y_1,y_2,\ldots)$ which is constructed solely out of the quark
operators in \seeq{quark}.

  Consider first a fixed background field $\cu\in\cg-\cg_0$. Suppressing
the coordinates $y_1,y_2,\ldots$ we find in the limit $N\to\infty$
\beq
   \lim_{N\to\infty} \vev{J^a_{5q}(x)\, O}_{\cu,N} =
   \bra{0_H} \hat{O}_L \co(m) \hat{O}_R \ket{0_H}
   \bra{0_H} \hat{j}^a_5(x) \ket{0_H} \,.
\label{fctr}
\eeq
The subscript $\cu$ indicates the we have given the expression for the
correlator in a fixed background. Here
\beq
   \hat{j}^a_5(x) =  \hat{c}_x\dg\l^a\hat{c}_x - \hat{d}_x\dg\l^a\hat{d}_x  \,.
\eeq
The operator expressions for $\hat{O}_L$ and $\hat{O}_R$ can
be found using the rules of App.~A.  For finite $N$, the difference between the
expressions on the \lhs and the \rhs of \seeq{fctr} obeys a bound analogous to
inequality~(6.3). The crucial observation is that the matrix element
$\sbra{0_H} \hat{j}^a_5(x) \sket{0_H}$ is proportional to $\tr\l^a$, and so
it vanishes identically for a non-singlet pseudo-scalar density. Hence
\beq
  \left| \vev{J^a_{5q}(x)\, O}_{\cu,N} \right| \le
  c\, 2^{2N_t} e^{-NE_0(\cu)} \,.
\label{j5bnd}
\eeq
The constant $c$ is the product of the operator norms
$\parallel\!\! \hat{O}_L\hat{O}_R \!\!\parallel$ and
$\parallel\!\! \hat{j}^a_5 \!\!\parallel$.
Integrating over the group variables and following the same reasoning as
before we conclude that
\beq
   \lim_{N\to\infty} \vev{J^a_{5q}(x)\, O}_N = 0 \,.
\eeq
This proves the restoration of NSAS in the {\it chiral limit} $N\to\infty$.
As an example, the explicit expression for the $N\to\infty$ limit of the
ward identity\seneq{pionms} is given in App.~A.

  We conclude this section with two comments. For the singlet current,
$\tr I \ne 0$, and the matrix element $\sbra{0_H} \hat{j}_5(x) \sket{0_H}$
is in general non-zero. As was shown in  refs.~\cite{gjk,anml,nn2,mcih}
the axial anomaly is reproduced correctly in the continuum limit.

  Our second comment concerns the factorized form of the \rhs of \seeq{fctr}.
In the limit $N\to\infty$, factorization of the expectation value of an
operator product occurs whenever the limiting $s$-separation between two
factors in the product tends to infinity. This means that correlation
functions obey clustering in the $s$-direction. Now, clustering is by itself
a rather weak condition, which we normally  expect  to hold even if massless
particles can propagate between two points.  As is often the case with rigorous
bounds, we believe that the actual damping of correlations in the $s$-direction
is much stronger, and that the correlation length in the $s$-direction is
finite in physical units, if not in lattice units. A heuristic
discussion of the actual magnitude of anomalous effects for finite $N$
is given in the concluding section.

\vspace{5ex}
\noindent {\large\bf 7.~~Conclusions}
\vspace{3ex}
\secteq{7}

  In this paper we discussed in detail a new formulation of lattice QCD which
is based on the surface fermions scheme. Our main result is that non-singlet
axial symmetries become exact in the chiral limit. The chiral limit
is defined to be the limit of an
infinite fifth direction, at fixed finite values of the bare coupling and
the size of the four dimensional lattice. The vanishing of the anomalous term
in all NSAS Ward identities implies in particular that non-singlet axial
currents do not undergo any renormalization in the chiral limit, as should be
the case for truely conserved currents.

  As was shown in previous works~\cite{gjk,anml,nn2,mcih},
the singlet axial anomaly is correctly
reproduced if one takes first the limit $N\to\infty$ and then the continuum
limit $g_0\to 0$. For finite $g_0$, the limiting ``$N=\infty$'' formulation is
therefore a new non-perturbative regulator of QCD which is maximally symmetric
under axial transformations. This new regularization scheme is mildly non-local
because we have integrated out an infinite number of heavy four-dimensional
fields. We believe that the mild non-locality does not jeopardize the
consistency of the continuum limit, but we have not investigated every possible
aspect of this issue.

  The non-singlet currents defined in \seeq{axcur} are exactly conserved in
the limit $N\to\infty$ regardless of the value of $g_0$. Thus, they will be
conserved not only in the continuum limit, but also in the strong coupling
limit. However, it is not necessarily true that in the strong coupling limit,
these currents retain the physical significance of {\it axial} current.  We do
not rule out that for sufficiently strong coupling, some or all of the doubler
modes reappear. The likely consequence would be that the currents
of \seeq{axcur} become {\it vectorial} with respect to the new massless
spectrum. In this case, the singlet current would become vectorial too, and
it will be conserved in the strong coupling phase.

  Returning to the physically interesting limit $g_0\ll 1$,
the biggest advantage of the new formulation is that fine tuning is
no longer needed in the fermion sector. For example, the theoretical
values of current masses are determined using weak coupling perturbation
theory, and they involve only multiplicative renormalization. Likewise,
operator mixing are restricted by the naive transformation properties
under non-singlet axial symmetries, and meson decay constants can be inferred
directly from the corresponding Ward identities.

  All the above properties become exact in the limit $N\to\infty$.
Looking forward to the implementation of the surface fermions scheme in
numerical simulations, it is important to have a realistic estimate of the
magnitude of anomalous effects on a finite five dimensional lattice.
While sufficient for proving the restoration of NSAS in the limit,
we believe that
the rigorous bounds used in Sect.~6 represent a gross overestimation
on the actual magnitude of anomalous effects. We have decide to
include here a short  heuristic discussion of this issue, mainly because we
believe that the true picture is much more promising. A more detailed study
is relegated to a separate publication.

  Our central observation is the following. As we noted previously, the naive
continuum limit of the lattice operator $D^\parallel$ is a massive Dirac
operator. This does not rule out the existence of (exact or approximate)
zero modes of $D^\parallel$
for certain gauge field configurations. But it does indicate that the relevant
configurations cannot in any sense be the discretized approximation of smooth
continuum gauge fields.  It may therefore be possible to prove the following
conjecture: if a given lattice gauge field is the discretized approximation of
a smooth continuum gauge field, then there is an $O(1)$ gap in the spectrum of
the (hermitian) operator $\g_5 D^\parallel$.
We have intentionally omitted here a precise definition of what we mean
by a ``lattice approximation of a smooth continuum gauge field''. Basically
what we have in mind is a condition which states that the gauge field's action
density is very small everywhere. However, there may be other definitions that
have the same physical content, and which are more convenient from a
mathematical point of view.

  The above conjecture receives circumstantial evidence from the two known
examples of zero modes of $D^\parallel$. These are the singular fluxon of
ref.~\cite{nn2}, and the singular dynamical domain wall of Sect.~5. Both are
characterize by an action density which is $O(1)$ on some $n$-dimensional
subspace, where $n=2$ and $n=3$ respectively for a fluxon and for a domain
wall.  Now, if we want  the configuration to support an approximate zero mode
whose energy is $E_0\ll 1$, then its longitudinal extension should be at least
$O(1/E_0)$. This implies that the total gauge field action for such
configurations should be bounded from below by $S_G\sim C/(g_0^2 E_0^n)$.
Here $C$ is some $O(1)$ constant.

  Typically, a single extended configuration supports a single (approximate)
zero mode, up to symmetry factors. In other words, level crossing in the
spectrum of $\ch$ occurs one at a time, and all other eigenstates
are separated by a finite gap in the vicinity of the crossing point. Assuming
this to be true in general, all factors of $2^{N_t}$ (which count the total
number of eigenstates) in the inequalities of Sect.~6 can be dropped. Putting
everything together, this suggests that the magnitude of the anomalous term for
a given background should be
\beq
\exp\{-C/(g_0^2 E_0^n) - N E_0\} \,.
\eeq

  To complete the estimate of the anomalous term, it is necessary to
determine what are the most important singular configurations.  If all
configurations have $n>0$, the most important effect should be the need to
increase the longitudinal extension of the singular configuration with
decreasing $E_0$. The dominant configurations, as well as the magnitude of the
resulting anomalous term, can then be determined by a saddle point
approximation. If, on the other hand, there exist point-like
four dimensional configurations which support (approximate) zero modes of
$\g_5 D^\parallel$, then a more refined analysis would be needed, and one
would have to estimate the phase space for such configurations as a function of
$g_0$.  We believe that, either way, the correlation length in the
$s$-direction should turn out to be finite. If it is found that, moreover,
the correlation length in the $s$-direction remains finite in
{\it lattice units}, then we may hope to obtain good results already on
five dimensional lattices which can be simulated today.

\vspace{3ex}
  This research was supported in part by the Basic Research Foundation
administered by the Israel Academy of Sciences and Humanities, and by a grant
from the United States -- Israel Binational Science Foundation.

\vspace{5ex}
\noindent {\large\bf A.~~Correlation functions in the $N=\infty$ limit}
\vspace{3ex}
\secteq{A}

  In this appendix we give the transfer matrix formulae for various
correlation functions.  The correlation functions for the quark
operators\seneq{quark} take a particularly simple form in the limit
$N=\infty$. For example, the expression for the quark condensate is
\beq
  \svev{\Bar{q} q}_\cu = -\bra{0_H} \hat{c}^\dagger \co(m) \hat{c} \ket{0_H}
                     -\bra{0_H} \hat{d}^\dagger \co(m) \hat{d} \ket{0_H} \,.
\eeq
The subscript $\cu$ indicates that we give the expression for the
correlator in a fixed background field.
Another example is the two pion correlator
\bqry
\label{pion}
   \vev{ \p^-(x) \p^+ (y)}_\cu & = &
   -\bra{0_H} \hat{c}_{x\uar}\dg \hat{c}_{y\dar}\dg \co(m)
              \hat{c}_{y\uar} \hat{c}_{x\dar} \ket{0_H}  \NON
& &+\bra{0_H} \hat{c}_{x\uar}\dg \hat{d}_{y\uar}\dg \co(m)
              \hat{d}_{y\dar} \hat{c}_{x\dar} \ket{0_H}  \NON
& &+\bra{0_H} \hat{d}_{x\dar}\dg \hat{c}_{y\dar}\dg \co(m)
              \hat{c}_{y\uar} \hat{d}_{x\uar} \ket{0_H}  \NON
& &-\bra{0_H} \hat{d}_{x\dar}\dg \hat{d}_{y\uar}\dg \co(m)
              \hat{d}_{y\dar} \hat{d}_{x\uar} \ket{0_H}\,.
\eqry
Here
\bqry
  \p^- & = & i \bar{q}_\uar \g_5 q_\dar \,,\NON
  \p^+ & = & i \bar{q}_\dar \g_5 q_\uar \,.
\eqry
The arrows denote isospin. Notice that the pion operators are special cases
of the pseudoscalar densities \seeq{psc}.

  The general prescription for correlation functions of the quark
operators \seeq{quark} is the following. Considering $q_{R,L}$ and
$\bar{q}_{R,L}$ as Grassmann variables, one first reorder each product of
quark operators such that $q_L$ and $\bar{q}_L$ occur to the left of all
$q_R$ and $\bar{q}_R$. This step may result in a minus sign. One then
translates the result into a matrix element of the form
\beq
  \bra{0_H} \cdots \co(m) \cdots \ket{0_H} \,.
\eeq
The operators to the left of $\co(m)$ are obtained from the ordered
product of $q_L$-s and $\bar{q}_L$-s by the substitution
\bqry
   \bar{q}_L & \to & \hat{c}\dg \,, \NON
         q_L & \to & \hat{d}\dg \,.
\eqry
Similarly, on the right of $\co(m)$ one makes the substitution
\bqry
   \bar{q}_R & \to & \hat{d} \,, \NON
         q_R & \to & -\hat{c} \,.
\eqry
All indices are left unchanged in this substitution. (The transition from
the Grassmann path integral to operator language involves a non-local
transformation at an intermediate step~\cite{l,nn2}. But this non-locality
cancels out in the final expression).

  Apart from quark operators, we may also be interested in the transfer matrix
formulae for correlation functions that involve vector or axial currents.
As an example, for finite $N$, the correlator on the \lhs of \seeq{wi} becomes
\beq
  \vev{A^a_\m(x)\, O(y_1,y_2,\ldots) }_\cu =
  \sum_{s=0}^{2N-1} \sign(N-s - \half)\,
  {\tr T^{2N} \hat{O}_L \co(m) \hat{O}_R\, \hat{j}^a_\m(x,s)
  \over  (\tr T^N)^2}
\eeq
where
\beq
   \hat{j}^a_\m(x,s) = T^s\, \hat{j}^a_\m(x)\, T^{-s} \,.
\eeq
To obtain an explicit expression for $\hat{j}^a_\m(x)$, the different
terms in \seeq{5dcrnt} are translated according to the following rules
\bqry
    \Bar\j_{L,x}\, \j_{R,y} & \to &
    - \hat{c}_y\, T\, \hat{c}\dg_x\, T^{-1} \NON
    \Bar\j_{R,x}\, \j_{L,y} & \to &
    - \hat{d}_x\, T\,  \hat{d}\dg_y\, T^{-1} \NON
    \Bar\j_{L,x}\, \j_{L,y} & \to &
    - T\, \hat{c}\dg_x\, \hat{d}\dg_y\, T^{-1} \NON
    \Bar\j_{R,x}\, \j_{R,y} & \to &
    - \hat{d}_x\, \hat{c}_y
\eqry
The somewhat unexpected appearance of the transfer matrix $T$ in these rules is
due to our definition of $\exp\{-H\}$ \seeq{exph}. If instead we decided to use
$\exp\{-H'\}=K\dg K$ and the corresponding transfer matrix $T'$, then there
would be no factors of $T'$ in resulting expression for $\hat{j}^a_\m(x)$.
On the other hand, the expressions for quark correlators would become more
cumbersome.

  Convergence of the infinite sum on the \lhs of every Ward identity is
guaranteed by the finiteness of the \rhs\ We comment that, if a slightly
stronger form of clustering than the one proved in Sect.~6 holds,
then
\bqry
   \lim_{N\to\infty} \vev{A^a_\m(x)\, O(y_1,y_2,\ldots) }_\cu =
   \sum_{s=0}^{\infty}
   \sbra{0_H} \hat{O}_L \co(m) \hat{O}_R\, \hat{j}^a_\m(x,s) \sket{0_H} \NON
   -\sum_{s=-\infty}^{-1}
   \sbra{0_H} \hat{j}^a_\m(x,s) \hat{O}_L \co(m) \hat{O}_R \sket{0_H} \,.
\eqry
This equation is in particular valid if, as we have every reason to believe,
the correlation length in the $s$-direction is finite.

\vspace{5ex}
\noindent {\large\bf B.~~Inequalities}
\vspace{3ex}
\secteq{B}

  \seEq{reflection} implies an analogous identity for fermion propagator,
considered as a matrix. Writing the coordinates explicitly one has
\beq
\label{prop}
\gamma_5 G(x,s;y,s') \gamma_5  = G\dg(y,2N+1-s';x,2N+1-s)\,.
\eeq
Here the dagger refers only to the the suppressed internal indices.
We will use this identity to prove the positivity of correlators of the
following pseudoscalar densities
\bqry
  K^a(x,t) & = & i\Bar\j(x,N+t) P_R\, \l^a \j(x,N+1-t) \NON
           & & -i\Bar\j(x,N+1-t) P_L\, \l^a \j(x,N+t) \,.
\eqry
Notice the special cases
\beq
\label{anm}
  K^a(x,0) = iJ^a_{5q}(x) \,,
\eeq
\beq
  K^a(x,N)= iJ^a_5(x) \,.
\eeq
Using \seeq{prop} and inserting a minus sign for the closed fermion loop
we now obtain
\beq
  \svev{K^a(x,t) K^b(y,t')}  = \d^{ab} Z^{-1} \int DU e^{-S(U)}
                                  \left(\det D_F(U)\right)^{N_f}
                                  I(U;x,t;y,t')\,,
\eeq
where
\bqry
\label{crl}
   I & = & \tr\left\{
       P_R\, G(y,N+t';x,N+1-t) P_R\, G\dg(y,N+t';x,N+1-t)  \right.  \NON
& &  + P_R\, G(y,N+1-t';x,N+1-t) P_L\, G\dg(y,N+1-t';x,N+1-t)       \NON
& &  + P_L\, G(y,N+t';x,N+t) P_R\, G\dg(y,N+t';x,N+t)               \NON
& &  + \left.
       P_L\, G(y,N+1-t';x,N+t) P_L\, G\dg(y,N+1-t';x,N+t) \right\}\,.
\eqry

  For even $N_f$ or for $m>0$, the factor $(\det D_F(U))^{N_f}$ is positive.
We assume that one of these conditions is satisfied.  It is now straightforward
to prove the positivity of each term in \seeq{crl}. Consider the second row as
an example. Independently of the values of $x$ and $y$, it has the generic form
\bqry
   \tr\left\{ P_R\, A\, P_L\, A\dg \right\} & = &
   \tr\left\{ P_R^2\, A\, P_L^2\, A\dg \right\} \NON
& = &  \tr\left\{ \left(P_R\, A\, P_L\right)
       \left(P_R\, A\, P_L\right)\dg \right\}\,.
\eqry
The last row is manifestly positive.

  Consider now the Ward identity\seneq{pionms}
which determines the pion mass. Notice that
\beq
  \p^a(y)=K^a(x,N) \,.
\eeq
Using \seeq{anm} the anomalous term in this Ward identity can be written as
\beq
  \svev{K^a(x,0) K^a(x,N)}\,,
\eeq
which is positive according to the previous discussion. The same reasoning
proves the positivity of the two-pion correlator.

  The positivity of the  two-pion correlator can also be established directly
from the $N=\infty$ formula \seeq{pion}. One should notice that the Fock
space is a direct product of the Up and Down Fock spaces. Thanks to
factorization of $\co(m)=\co_\uar(m)\,\co_\dar(m)$, each
matrix element in \seeq{pion} becomes the product of two complex conjugate
matrix elements, one in the Up Fock space and one in the Down Fock space.

\vspace{5ex}
\centerline{\rule{5cm}{.3mm}}




\begin{thebibliography}{99}

\bibitem{an} J. Bell and R. Jackiw, Nouv. Cim. {\bf 60A} (1969) 47.
S. Adler, \PR{177} (1969) 2426.

\bibitem{ab} S. Adler and W. Bardeen, \PR{182} (1969) 1517

\bibitem{rg} A. Zee, \PRL{29} (1972) 1198.
S.Y. Pi and S.S. Shei, \PRD{11} (1975) 2946.

\bibitem{mb} M. Bos, \NPB{404} (1993) 215.

\bibitem{w} Wilson  K.G.~Wilson, in {\it New Phenomena in Sub-Nuclear Physics}
(Erice, 1975), ed. A.~Zichichi (Plenum, New York, 1977).

\bibitem{ks} L.H.~Karsten and J.~Smit, \NPB{183} (1981) 103.

\bibitem{it} M. Bochicchio, L. Maiani, G. Martinelli, G.C. Rossi and M. Testa,
\NPB{262} (1985) 331. C. Curci, \PLB{167} (1986) 425.

\bibitem{ca} S. Treiman, R. Jackiw, B. Zumino and E. Witten, {\it Current
Algebra and Anomalies}, World Scientific, 1985.

\bibitem{weak} N. Cabibbo, G. Martinelli and R. Petronzio, \NPB{244} (1984)
381. R.C. Brower, M.B. Gavela, R. Gupta and G. Maturana, \PRL{53} (1984) 1318.
C. Bernard, T. Draper, G. Hockney, A.M. Rushton and A. Soni, \PRL{55} (1985)
2770. C. Bernard, A. Soni and T. Draper, \PRD{36} (1987) 3224.

\bibitem{weakstg} G.W. Kilcup and S.R. Sharpe, \NPB{283} (1987) 493.
S.R. Sharpe, A. Patel, R.Gupta, G. Guralnik and G.W. Kilcup,
\NPB{286} (1987) 253.

\bibitem{sf} S.A. Frolov and A.A. Slavnov, \NPB{411} (1994) 647.

\bibitem{k} D.B.~Kaplan, \PLB{288} (1992) 342; \NPBP{30} (1993) 597.

\bibitem{nn1} R. Narayanan and H. Neuberger, \PLB{302} (1993) 62.

\bibitem{bndr} Y. Shamir, \NPB{406} (1993) 90.

\bibitem{mcih} M. Creutz and I. Horvath, hep-lat/9312068, talk at LATTICE 93,
to appear in the proceedings; BNL preprint 60062, hep-lat/9402013.

\bibitem{nn2} R. Narayanan and H. Neuberger, \PRL{71} (1993) 3251;
\NPB{412} (1994) 574; Rutgers preprint RU-93-52, to appear in the proceedings
of LATTICE 93.

\bibitem{nogo} Y. Shamir, \PRL{71} (1993) 2691.

\bibitem{anml} Y. Shamir, Weizmann preprint WIS-93/99, hep-lat/9310006,
to appear in \NPB.

\bibitem{gjpv} M.F.L.~Golterman, K.~Jansen, D.N.~Petcher and J.C.~Vink,
preprint UCSD/PTH 93--28, Wash.~U. HEP/93--60, hep-lat/9309015.

\bibitem{gjk} M.F.L.~Golterman, K.~Jansen and D.B.~Kaplan, \PLB{301} (1993)
219.

\bibitem{jsc} K.~Jansen and M.~Schmaltz, \PLB{296} (1992) 374.
K. Jansen, \PLB{288} (1992) 348.

\bibitem{nnp} H.B.~Nielsen and M.~Ninomiya, \NPB{185} (1981) 20,
{\it Erratum} \NPB{195} (1982) 541; \NPB{193} (1981) 173.
A.~Pelissetto, \APH{182} (1988) 177.

\bibitem{l} M. L\"uscher, \CMP{54} (1977) 283.

\bibitem {susy} A. Casher and Y. Shamir, \PRD{39} (1989) 514.

\bibitem{thooft} G. `tHooft, \PRD{14} (1976) 3432.

\bibitem{vw} C.Vafa and E.Witten  \NPB{234} (1984) 173.

\bibitem{ch}  C.G.~Callan and J.A.~Harvey, \NPB{250} (1985) 427.



\end{thebibliography}
\end{document}